# Periods Detected During Analysis of Radioactivity Measurements Data

A.G.Parkhomov
Institute for Time Nature Explorations.
Lomonosov Moscow State University, Moscow, Russia.
alexparh@mail.ru

Analysis results of data of long-term radioactivity measurements of $^{3}H$, $^{56}Mn$, $^{32}Si$, $^{36}Cl$, $^{60}Co$, $^{137}Cs$, $^{90}Sr$-$^{90}Y$, $^{226}Ra$, $^{238}Pu$ and $^{239}Pu$ sources are presented. For beta-radioactive sources, their activity in addition to the exponential drop is characterized by rhythmic variations with a period of 1 year and magnitude of 0.1-0.35% from the average value. These oscillations attain maximum values between January and March, with corresponding minimum values occurring from July to September. Spectral analysis of 7-year long measurements of count rates of beta-radioactive sources $^{90}Sr$-$^{90}Y$ as well as 15-year long measurements of beta-gamma radiation of $^{226}Ra$ revealed presence of rhythmic variations with a period of about a month and magnitude 0.01%. Magnitude of diurnal oscillations did not exceed 0.003%. Analysis of measurements data for alpha radioactive sources $^{238}Pu$ and $^{239}Pu$ did not reveal any statistically reliable periodic patterns. Possible factors underlying these rhythmic oscillations in beta-radioactivity are discussed.

In recent years, long-term measurements of beta-decay produced evidence of periodic variations in the activity of various sources, in addition to the expected exponential decrease. Such variations were detected while registering beta-decays of $^{3}H$, $^{56}Mn$, $^{32}Si$, $^{36}Cl$, $^{60}Co$, $^{137}Cs$, $^{90}Sr$-$^{90}Y$, as well as those of decay products of $^{226}Ra$, but were not found in data from alpha-radioactive sources $^{238}Pu$ and $^{239}Pu$ (cf. table).

**Table.** Results of long-term measurements of radioactivity.

| Ref. | Sources | Decay type | Detector | Time span | Periods, days (magnitude, %) | Months of high/low values |
|---|---|---|---|---|---|---|
| 3 | $^{56}Mn$ | $\beta^-$ | NaJ+PMT | 02.78-01.87 | 365(0,3) | 01/07 |
| 1 | $^{3}H$ | $\beta^-$ | Scint.+photodiode | 11.80-05.82 | 365(0,35) | 02/08 |
| 4,14 | $^{32}Si$ | $\beta^-$ | Prop. Counter | 02.82-02.86 | 365(0,12); 32,5(0,01) | 02/08 |
| 4 | $^{36}Cl$ | $\beta^-$, e.c. | Prop. Counter | 02.82-02.86 | 365(0,12) | 02/08 |
| 2,14 | $^{226}Ra$ | $\alpha$, $\beta^-$ | Ion. Chamber | 10.83-06.99 | 365(0,1); cf. fig.3 | 01/08 |
| 9 | $^{137}Cs$ | $\beta^-$ | NaJ+PMT | (19-23)04.94 | 1(0,06)? | |
| 8 | $^{238}Pu$ | $\alpha$ | Energy evolved | 10.97-10.99 | None | |
| 10 | $^{137}Cs$, $^{60}Co$ | $\beta^-$, $\beta^-$ | NaJ+PMT | 12.98-04.99 | 1(?); ~30 | |
| 5-7 | $^{60}Co$ | $\beta^-$ | G-M counter | 03.99-10.03 | 365(0,2) | 03/09 |
| 11,12 | $^{137}Cs$, $^{60}Co$ | $\beta^-$, $\beta^-$ | Ge(Li) semic. | 03.00-04.00 | 1(0,5)? | |
| 5-7 | $^{90}Sr$-$^{90}Y$ | $\beta^-$, $\beta^-$ | G-M counter | 01.00-12.10 | 365(0,13); cf. fig.1 | 03/09 |
| 5-7 | $^{239}Pu$ | $\alpha$ | Si-semic. | 02.00-05.03 | None | |
| 5-7 | $^{90}Sr$-$^{90}Y$ | $\beta^-$, $\beta^-$ | G-M counter | 10.02-12.10 | 365(0,15) | 03/09 |
| 6 | $^{239}Pu$ | $\alpha$ | Si-semic. | 10.06-12.10 | None | |

It should be noted that rhythmic variations in radioactivity do not follow from the existing theory of this phenomenon, while their detection represents an experimental challenge, which requires fixed conductions of measurements with stable equipment running without interruptions for many years. It is no wonder that this effect has been discovered only recently, although the phenomenon of radioactivity has been known for more than 100 years. In most cases it manifested itself as an unidentified error in measurements of half-lives of long-lived radionuclides. Experiments described in [1, 5, 6] are different in that they were conducted with the aim of detecting deviations from the exponential decay law. They were carried out on specially created experimental installations.

**Rhythms of beta-radioactivity with a period of 1 year**

The **table** includes 7 different radionuclides which exhibited rhythmic variations with a period of 1 year, registered by using detectors of 5 different types. In all cases, the magnitude of the annual rhythm ranges between 0.1 and 0.35% from average count rate, with maximum values taking place between January and March, and minimum values from July to September. Close agreement of the results obtained with different radionuclides and by using different experimental methods, enables to state with confidence that the underlying cause of the periodic patterns in the count rates are the variations in the rates of radioactive decays rather than drifts and non-stabilities of registering equipment.

**Rhythms of radioactivity with a period of about a month**

Large amount of data accumulated in a number of experiments allows to apply spectral analysis which enables not only to pinpoint parameters of the observed annual rhythms but also to uncover other periodic patterns that remain unnoticed against the background of fluctuations and noise acting at random instants.

Existence of variations modulating radioactive decays with a monthly period (gamma radiation from $^{137}Cs$, $^{60}Co$) was reported for the first time in [10]. This publication does not make it clear how large these variations are, and the duration of the measurements (4 months) is not sufficient to determine the period rather exactly.

Results of analysis of significantly longer measurements are presented in [5-7]. Fig. 1 shows amplitude spectrum (a periodogram) of count rate measurement results for beta particles emitted by $^{90}Sr$-$^{90}Y$ source and registered by G-M counter. Analysis was performed by means of Fast Fourier Transform with subsequent conversion of frequency to periods.

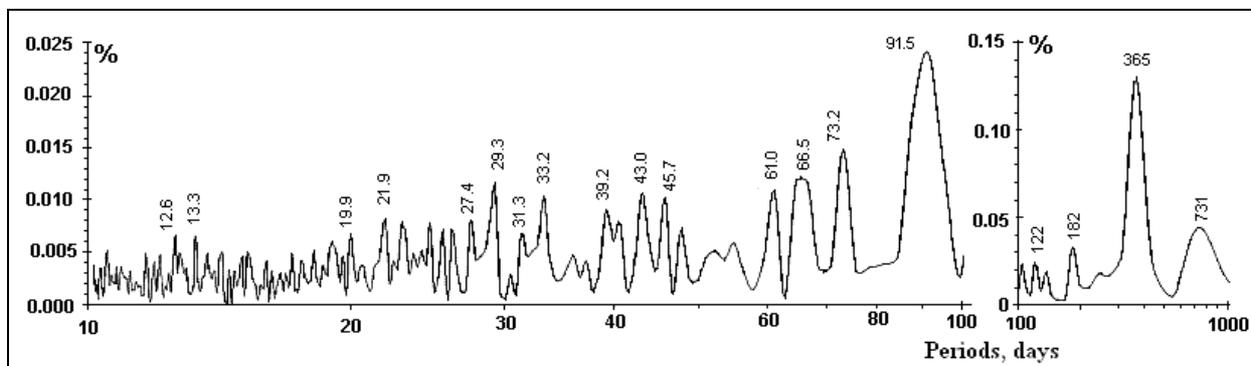

**Fig. 1.** Periodogram of variations in the count rate of $^{90}Sr$-$^{90}Y$ beta source as registered by G-M counter. The experiment was run between April 2000 and March 2007. Magnitude is shown as percentage of the average count rate [6,7]. Numbers correspond to the periods of nearby peaks.

Peak with the period of 1 year (with magnitude of 0.13%) and its harmonics (182, 91.5, 61 days) stand out. In the near-month range of periods, peaks with magnitude of 0.01% corresponding to 29.3 and 33.2 days dominate the landscape. A group of peaks with periods of about 43 days are readily discernible. Persistent nature of variations with the period of the synodical Lunar month (29.53 days) is demonstrated by averaging data of radioactivity measurements of a $^{90}Sr$-$^{90}Y$ source over 87 cycles (fig. 2).

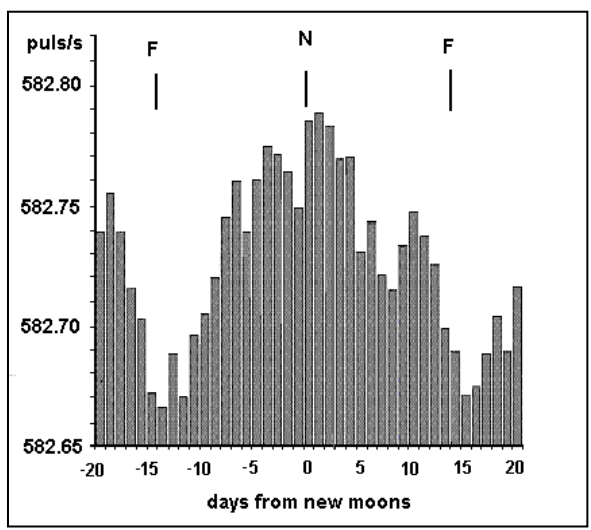

**Fig. 2.** Averaging over cycles of the synodic Lunar month of the count rate data for $^{90}Sr$-$^{90}Y$ beta source, as registered by G-M counter between April 2000 and March 2007. Averaging covers 87 cycles. N – new Moon, F - full Moon [6,7].

In [14], spectral analysis findings based on results of several years of radioactivity measurements of $^{226}Ra$ [2] and $^{32}Si$ [4] are presented. In addition to an annual period, this analysis reveals many other peaks in the frequency range from 4 to 20 year$^{-1}$ (corresponding to periods from 91 to 18 days) (fig. 3). The most pronounced of the near-month peaks are found to be 33.2, 32.3 and 30.2 days long. A group of peaks with periods about 43 days also stand out.

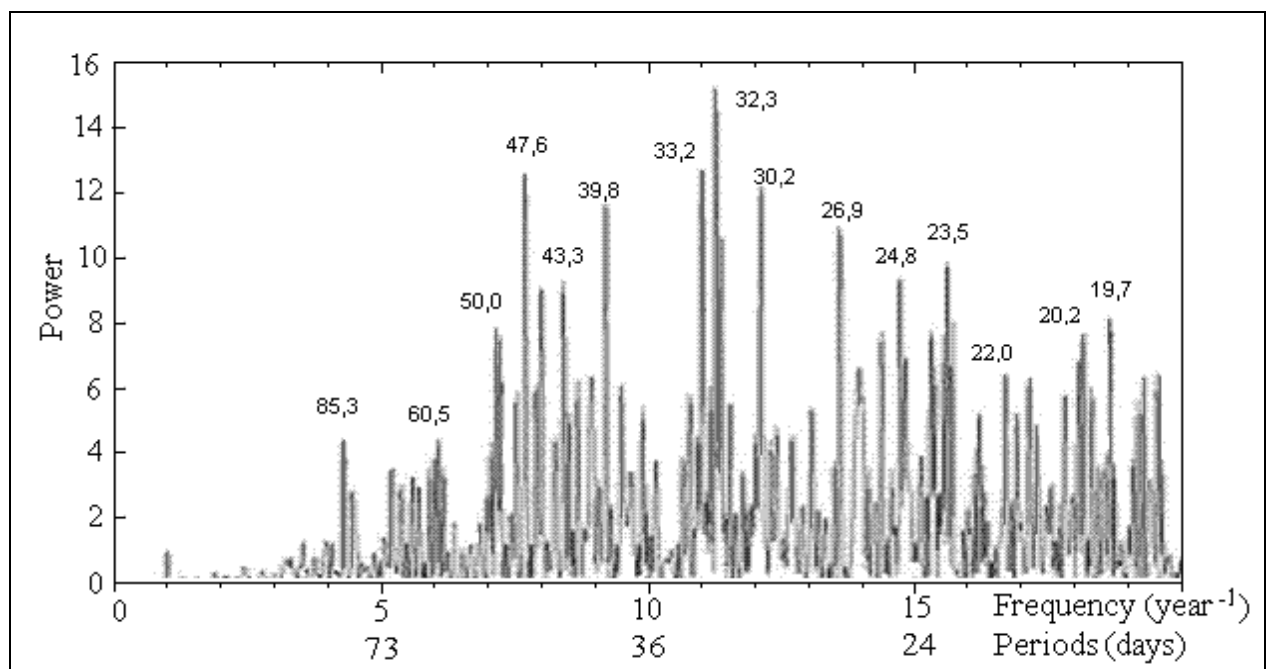

**Fig. 3.** Power spectrum of variations in the intensity of beta and gamma radiation from the $^{226}Ra$ source, as registered by ionization chamber. The analysis corresponds to time interval between November 1983 and October 1999 (1966 measurements in total) [14]. Numbers next to peaks specify their corresponding periods.

Collating fig. 2 with fig. 1 allows one to arrive at conclusion about good agreement between positions of many peaks, including those with near-month periods, calculated from these independent measurements, followed by different data processing schemes.

**Rhythms of radioactivity with diurnal periods**

The first report of rhythmic variations with a diurnal period and magnitude of about 0.06% in count rates for a beta radionuclide is contained in [9]. These oscillations were detected in measurements of gamma radiation from $^{137}Cs$ source with total duration of 4 days. Existence of a diurnal rhythm in results of analogous measurements was pointed out in [10], but the data in this paper cannot be used to estimate magnitude of the effect.

Count rate bursts with a diurnal period and magnitude of up to 1% were detected while registering gamma radiation from $^{60}Co$ and $^{137}Cs$ sources by means of Ge(Li) detector [11,12].

A diurnal modulation was detected by conducting spectral analysis of results of 7-year long count rate measurements for beta particles emitted by $^{90}Sr$-$^{90}Y$ source and registered by G-M counter [5-7] (fig. 4).

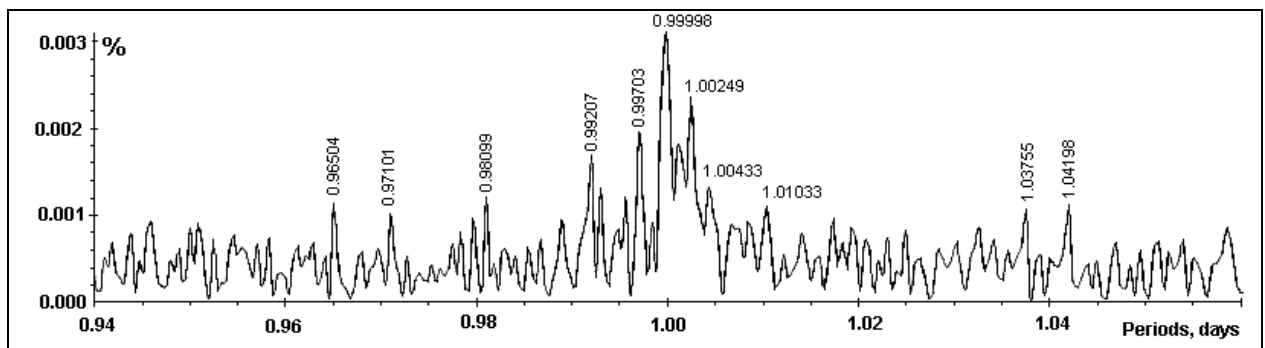

Fig. 4. Periodogram of variations in count rates for $^{90}Sr$-$^{90}Y$ beta source, as registered by G-M counter, for periods in the region around 1 day [6,7]. Time interval spans April 2000 and March 2007. Magnitude is shown as percentage value of the average count rate. Numbers next to peaks specify their corresponding periods.

In the region of near-day periods, the peak of the Solar day is clearly visible along with a fine structure produced by interaction of this rhythm with the annual rhythm and its harmonics. The peak of the Lunar day (1.03755) is also discernible. It is however possible that it corresponds to a combination frequency of the Solar day and the Lunar month, rather than reflecting some influences with a period of the Lunar day. This interpretation is suggested by the presence of the symmetric (in frequency domain) peak at 0.96504. Magnitude of diurnal variations does not exceed thousandths of percent of the average value and, as opposed to the modulations with yearly and monthly timescales, one could not affirm with confidence that they are not a result of thermally caused drifts in the measuring equipment, since averaged daily behavior of count rates (lowest at night, highest around midday) resembles diurnal oscillations of temperature.

A significantly higher value of modulation amplitude with a period of 1 day described in [9,11,12] is in contradiction with the results of [5-7]. One can assume that despite the precautions taken, experiments [9,11,12] were not completely free from being influenced by the oscillations of temperature in the vicinity of the installation which resulted in such large deviations. This conclusion is suggested by coincidence of the days on which no apparent effect was found with overcast weather conditions when temperature variations are generally low.

The issue of the range of diurnal oscillations of radioactivity, if they do indeed exist, remains thus open. Their magnitude probably does not exceed one thousandth of percent.

**Discussion**

Let us consider how the discovered properties of this phenomenon square with hypotheses put forward as an explanation for deviations from the purely exponential character of radioactive decay.

1. It was suggested that rhythmic variations in radioactive decays are associated with the changes in orientation relative to a global anisotropy of physical space [12]. But if the effect under consideration depends only on orientation relative to some direction, then modulations would occur with a period of the sidereal day and constant magnitude during the year. Experimental results show that, on the contrary, the most pronounced deviations take place with a yearly period, while diurnal oscillations are far weaker. Moreover, this hypothesis does not explain the fact that rhythmic variations are observed only for beta, and not alpha radioactivity.

2. A reasonable explanation of the fact that the effect under discussion occurs in beta decays only is to assume that it is caused by an incoming flux of cosmic neutrinos, since these particles are involved in processes of beta decay but do not take part in alpha decays. The hypothesis that the neutrino flux is produced by the Sun and that the annual changes in radioactivity are associated with the variations in the flux density as a result of changing distance between the Sun and the Earth due to the orbital motion of our planet, was proposed by Falkenberg [1] and reproduced by Jenkings and Fishbach [15,16].

This assumption looks very unconvincingly due to the extremely weak interactions with matter of solar neutrinos with energies of 1 MeV and higher. This elusive nature of neutrinos was confirmed by many experiments. If one takes that for some reason interaction of solar neutrinos with radionuclides is significantly stronger than suggested by currently accepted estimates, then their flux density would experience noticeable attenuation due to passing through the body of the Earth and result in a lower rates of radioactive decays at night when compared to the midday values. Experiments reveal no such changes. In addition, this hypothesis is not able to explain the existence of readily discernible variations with the period of the synodical Lunar month with maximums around new Moons and minimums around full Moons (cf. fig. 2).

Another hypothesis proposes that periodic changes in beta radioactivity are a manifestation of fluxes of relic neutrinos (more precisely, of one of the components of dark matter - slow neutrinos with velocity of 10…1000 *km/s*) [6,7,17,18]. This assumption is in agreement with the significantly lower magnitude of diurnal oscillations as compared to the oscillations with an annual rhythm. As shown in [18], the strength of the effect depends strongly on the velocity of motion relative to the flux of slow neutrinos. Throughout the year, due to the orbital motion of the Earth, its speed relative to Galactic neutrinos changes by 40 *km/s*, whereas the speed changes caused by the spinning of the Earth around its axis do not exceed 1 *km/s*.

The appearance of the rhythm of the syndic Lunar month (about 29.5 days) could be explained by the fact that the gravitational field in the system Earth-Moon-Sun changes with this period. And this gravitational field is the main factor affecting motion of the fluxes of slow neutrinos. Moreover, the assumption of the influence exerted by cosmic fluxes of slow neutrinos on beta radioactivity explains bursts of the radioactivity of beta sources mounted in the focal point of a parabolic mirror [7, 18, 19].

The origin of the rhythm with period of 33 days is unknown. This timescale is not only distinctly visible on spectral curves of variations in radioactivity of $^{90}Sr$-$^{90}Y$ and $^{226}Ra$ sources but also manifests itself in spectral analysis of the results of 18-year long measurements of gravitational constant [20] on installation with torsion balance [21]. This period differs in a reliable way from the lunar rhythms and rhythms of the solar activity. An attempt to associate the 33-day period with the rotation of the Sun's core [14] does not seem convincing.

The author is deeply indebted to V.V. Kapelko for actual support of my researches and translation of my papers into English. The author is grateful to E.A. Kramer-Ageev, E.F. Makliaev, N.F. Perevozchikov, B.M. Vladimirsky, A.P. Levich, S.E. Shnoll, V.I. Muromtsev, A.V. Karavaikin, N.V. Samsonenko, D.S. Baranov, Yu.A. Baurov, L.B. Boldyreva for support, invariable attention to these researches and helpful conversations.

———